\documentclass[twocolumn,amsmath,amssymb,prl,preprintnumbers]{revtex4}
\usepackage{graphicx}% Include figure files
\usepackage{dcolumn}% Align table columns on decimal point
\usepackage{bm}% bold math
\usepackage{amsmath,amssymb}
\usepackage{color}
\usepackage{longtable}
\usepackage{ulem}
\usepackage{longtable} 
\usepackage{ulem}

\newcommand{\be}{\begin{equation}}
\newcommand{\ee}{\end{equation}}
\newcommand{\bea}{\begin{eqnarray}}
\newcommand{\eea}{\end{eqnarray}}
\newcommand{\bi}{\begin{itemize}}
\newcommand{\ei}{\end{itemize}}

\newcommand\QQbar{Q\overline{Q}}
\newcommand\mQ{m_{Q}}
\newcommand\SdotS{{\mathbf S}_Q\cdot{\mathbf S}_{\overline{Q}}}

\newcommand\SQ{{\mathbf S}_Q}
\newcommand\SQbar{{\mathbf S}_{\overline{Q}}}
\newcommand\tsrc{t_{\rm s}}

\begin{document}

%-- title ---
\title{
Charmonium potential from full lattice QCD
}

%-- author list ---

\author{Taichi Kawanai${}^{1,2,3}$}
\email{kawanai@nt.phys.s.u-tokyo.ac.jp}
\author{Shoichi Sasaki${}^{1}$}
\email{ssasaki@phys.s.u-tokyo.ac.jp}
\affiliation{${}^{1}$Department of Physics, The University of Tokyo, Hongo 7-3-1, Tokyo 113-0033, Japan}
\affiliation{${}^{2}$RIKEN-BNL Research Center, Brookhaven National Laboratory
Upton, NY 11973-5000, USA}
\affiliation{${}^{3}$Department of Physics, Brookhaven National Laboratory
Upton, NY 11973-5000, USA}

%-- preprint number ---
\preprint{TKYNT-11-09}

\date{\today}

%-- abstract ---
\begin{abstract}
 We present both spin-independent and -dependent parts of a central interquark potential 
 for charmonium states, which is  calculated
 in 2+1 flavor dynamical lattice QCD using the PACS-CS gauge configurations with a lattice cutoff 
 of $a^{-1}\approx 2.2$ GeV. Our simulations are performed with a relativistic heavy-quark action for the 
 charm quark at the lightest pion mass, $M_\pi=156(7)$ MeV,  
 in a spatial volume of $(3 \;\text{fm})^3$. 
 We observe that the spin-independent charmonium potential obtained from lattice QCD with almost 
 physical quark masses is quite similar to the Cornell potential 
 used in nonrelativistic potential models.
  The spin-spin potential, which is calculated in full lattice QCD for the first time, 
  properly exhibits a finite-range repulsive interaction. Its $r$-dependence is different
  from the Fermi-Breit type potential, which is widely
  adopted in quark potential models.
%, not point-like contact interaction 
%known as  Fermi-Breit type interaction.}
% The spin-spin potential properly exhibits the short range repulsive interaction, while its $r$-dependence
% is different from either a point-like spin-spin potential generated by one-gluon exchange or
% a phenomenological finite-range one adopted in quark potential models.

\end{abstract}


\maketitle

%--- main text ---------------------------------------  ----------------

%
%intro
%
Recently, many of the newly discovered charmoniumlike mesons
have been announced by $B$-factories at KEK and SLAC, which
are primarily devoted to the physics of CP violation.
Such states named as ``$XYZ$'' mesons could not be simply explained 
by a constituent quark description in quark potential 
models~\cite{Voloshin:2007dx}.
Thus, the charmoniumlike $XYZ$ mesons are expected to be good 
candidates for nonstandard quarkonium mesons 
such as hadronic molecular states, diquark-antidiquark bound 
states (tetraquark states) or hybrid mesons~\cite{Godfrey:2008nc}.
However, it seems to be still too early to judge whether we may discard the 
constituent quark description for such higher-mass charmonium states.

The interquark potential in quark potential models is
based on the phenomenology of confining quark interactions: the
so-called Cornell potential~\cite{Eichten:1975ag} together with spin-spin, 
tensor and spin-orbit terms appeared as leading spin-dependent corrections 
in powers of the  inverse of the heavy-quark mass
$\mQ$~\cite{Barnes:2005pb,Godfrey:1985xj}. Although the Cornell potential
was qualitatively justified by the static heavy-quark potential 
obtained from Wilson loops in lattice QCD~\cite{Bali:2000gf},  
the functional forms of the spin-dependent terms in potential models
are basically determined by perturbative one-gluon exchange as 
the Fermi-Breit type potential~\cite{Godfrey:1985xj}.
Thus, properties of higher-mass charmonium states predicated 
in potential models may suffer from large uncertainties, 
since the phenomenological spin-dependent potentials 
would have validity only at short distances and also 
in the vicinity of the heavy-quark mass limit.

In this sense, the reliable interquark potential directly derived 
from first principles QCD is desired at the charm quark mass. 
Indeed, the static potential between infinitely heavy-quark 
and antiquark, which is obtained from Wilson loops, 
have been precisely calculated in lattice QCD simulations~\cite{Bali:2000gf}.
The relativistic corrections to the static potential are 
classified in powers of $1/\mQ$ within a framework called potential 
nonrelativistic QCD~\cite{Brambilla:2004jw}.
The leading and next-to-leading order corrections
have been successfully calculated in quenched lattice QCD
with high accuracy by using multilevel algorithm~\cite{Koma:2006fw}.
However, it is rather difficult to calculate the proper {\it charmonium} 
potential in lattice QCD within the Wilson loop formalism. 
It is obvious that the inverse of the charm quark mass is 
far outside the validity region of the $1/\mQ$ expansion. 
Indeed, a spin-spin potential determined at ${\cal O}(1/\mQ^2)$~\cite{Koma:2006fw},
which exhibit an attractive interaction for the higher spin states,
seems to yield wrong mass ordering among hyperfine multiplets.
In addition, practically, the multilevel algorithm employed 
in Ref.~\cite{Koma:2006fw} is not easy
to be implemented in dynamical lattice QCD simulations. 

Under these circumstances, in our previous work~\cite{Kawanai:2011xb}, we have 
proposed a novel approach, where the interquark potential at finite quark mass can 
be accurately determined  from the equal-time and Coulomb gauge Bethe-Salpeter (BS) 
amplitude through an effective Schr\"odinger equation (See also Ref.~\cite{Ikeda:2011bs}).
The BS amplitude method is originally applied for the hadron-hadron interaction 
~\cite{{Ishii:2006ec}}. As demonstrated in Ref.~\cite{Kawanai:2011xb}, 
the spin-independent part of an interquark potential calculated in the new method smoothly approaches
the static potential given by Wilson loops in the infinitely heavy-quark limit.
The new approach enables us to determine both spin-independent and spin-dependent
interquark potentials at the charm quark mass, which potentially account for all orders of
$1/\mQ$ corrections. Furthermore, there is no restriction to dynamical calculation 
within this method.

In this paper, we present results of the spin-independent 
central and spin-spin potentials between the quark~($Q$) and 
antiquark~($\overline{Q}$) at the vicinity of the physical charm 
quark mass from the BS amplitude with 2+1 flavor PACS-CS gauge 
configurations~\cite{Aoki:2008sm}, where the simulated 
pion mass is closest to the physical point as $M_\pi=156(7)$  MeV.
As for a treatment of heavy quarks, we adopt a relativistic heavy 
quark (RHQ) action, which can remove large discretization errors 
introduced by large quark mass~\cite{Aoki:2001ra}.
%
% table 1
%
\begingroup
\begin{table*}%[H]
  \caption{Summary of RHQ parameters and results of  $S$- and $P$-wave charmonium masses.
      \label{Tab2}
      }
      \begin{ruledtabular}
      \begin{tabular}{ccccccccc} %\hline  \hline 
       RHQ parameters &$\kappa_c$ & $\nu$ & $r_s$  & $c_B$ & $c_E$ & \\ \hline
       & 0.10819 & 1.2153 & 1.2131  & 2.0268 & 1.7911 & \\  \hline \hline
       Charmonium mass& $M_{\text{ave}}^{S\text{-wave}}$ & $M_{\text{hyp}}^{S\text{-wave}}$ &
       $\eta_c$ ($0^{-+}$) & $J/\psi$ ($1^{-+}$) & 
       $\chi_{c0}$ ($0^{++}$) & $\chi_{c1}$ ($1^{++}$) & $h_{c}$ ($1^{+-}$)\\ \hline
%       [GeV]& 3.0702(9) & 0.1137(8) & 2.9849(4) & 3.0987(11) & 3.3925(60) &  3.4862(60) &
       [GeV]& 3.0638(9) & 0.1124(9) & 2.9794(5) & 3.0919(10) & 3.3865(58) &  3.4781(62) & 3.4995(62)
      \end{tabular}
      \end{ruledtabular}
\end{table*}
\endgroup
%
%

%%%%%%%%%%%%%%%%%%%%%%%%%%%%%%%%%%%%%%%%%%%%%%%%%%%%%%%%
% formulation
%%%%%%%%%%%%%%%%%%%%%%%%%%%%%%%%%%%%%%%%%%%%%%%%%%%%%%%%
Let us briefly review the new method utilized here to
calculate the interquark potential with the  finite quark mass.
As the first step, we consider the following equal-time $\QQbar$ 
BS amplitude in the Coulomb gauge for the 
quarkonium states~\cite{{Velikson:1984qw},{Gupta:1993vp}}:
\begin{equation}
  \phi_\Gamma({\bf r})= \sum_{{\bf x}}\langle 0| \overline{Q}
  ({\bf x})\Gamma Q({\bf x}+{\bf r})|
  \QQbar;J^{PC}\rangle \label{eq_phi},
\end{equation}
where ${\bf r}$ is the relative coordinate of two quarks at certain
time slice and $\Gamma$ is any of the 16 Dirac $\gamma$ matrices. 
A summation over spatial coordinates ${\bf x}$ projects onto 
the zero total momentum. The ${\bf r}$-dependent amplitude, 
$\phi_\Gamma({\bf r})$, is here called BS wave function.

In lattice simulations, the BS wave function can be extracted from 
the following four-point correlation function 
\begin{eqnarray}
  && \sum_{{\bf x},{\bf x}^{\prime}, {\bf y}^{\prime}}
  \langle 0  |\overline{Q}({\bf x}, t)\Gamma Q({\bf x}+{\bf r}, t) 
  \left(\overline{Q}({\bf x}^{\prime},\tsrc)
  \Gamma Q({\bf y}^{\prime},\tsrc) \right)^{\dagger}
  |0\rangle\nonumber \\
  &&=   \sum_{{\bf x}}\sum_{n} A_n\langle 0|\overline{Q}({\bf x})\Gamma Q({\bf x}+{\bf r}) 
  |n \rangle e^{-M^\Gamma_n(t-\tsrc)}\label{eq_correlator} \nonumber\\
  &&\xrightarrow{t\gg \tsrc} A_0 \phi_\Gamma({\bf r})  
  e^{-M^\Gamma_0(t-\tsrc)},
\end{eqnarray}
at the large Euclidean time from source location ($\tsrc$).
Here both quark and antiquark fields at $\tsrc$ are
separately averaged in space as wall sources.
The constant amplitude $A_n$ is a matrix element defined as 
$A_n=\sum_{{\bf x}^{\prime}, {\bf y}^{\prime}}
\langle n |\left(\overline{Q}({\bf x}^{\prime})
\Gamma Q({\bf y}^{\prime}) \right)^{\dagger}|0\rangle$.
$M^\Gamma_n$ denotes a mass of the $n$-th quarkonium state $|n\rangle$ in
a given $J^{PC}$ channel.  For instance, when $\Gamma$ is chosen  
to be $\gamma_5$ for the pseudoscalar (PS) channel $(J^{PC}=0^{-+})$ 
and $\gamma_i$ for the vector (V) channel $(J^{PC}=1^{--})$  
in the charm sector, $M_0^{\text{PS}}$ and $M_0^{\text{V}}$ correspond to 
the rest masses of the $\eta_c$ and $J/\psi$ ground states,
which can be read off from  the asymptotic large-time behavior 
of the correlation functions.

The BS wave function satisfies an effective Schr\"odinger
equation with a nonlocal and energy-independent interquark potential 
$U$~\cite{Ishii:2006ec,Caswell:1978mt} 
\begin{equation}
  -\frac{\nabla^2}{2\mu}\phi_\Gamma({\bf r})+
  \int dr'U({\bf r},{\bf r}')\phi_\Gamma({\bf r}')
  =E_\Gamma\phi_\Gamma({\bf r}),
  \label{Eq_schr}
\end{equation}
where the reduced mass $\mu$ of the $\QQbar$ system is given by $\mQ/2$ 
with the quark kinetic mass $\mQ$. The energy eigenvalue $E_\Gamma$ of 
the stationary Schr\"odinger equation is supposed to be $M_\Gamma-2\mQ$~\cite{footnote1}.
If the relative quark velocity $v=|{\nabla}/\mQ|$ is small as $v \ll 1$, 
the nonlocal potential $U$ can generally expand 
in terms of the velocity $v$ as 
$U({\bf r}',{\bf r})=
 \{V(r)+V_{\text{S}}(r)\SdotS+V_{\text{T}}(r)S_{12}+
 V_{\text{LS}}(r){\bf L}\cdot{\bf S} + \mathcal{O}(v^2)\}\delta({\bf r}'-{\bf r})$,
 where $S_{12}=(\SQ\cdot\hat{r})(\SQbar\cdot\hat{r})-\SdotS/3$
 with $\hat{r}={\bf r}/r$, ${\bf S}=\SQ+\SQbar$
 and ${\bf L} = {\bf r}\times (-i\nabla)$~\cite{Ishii:2006ec}.
 Here, $V$, $V_{\text{S}}$, $V_{\text{T}}$ and $V_{\text{LS}}$ represent
 the spin-independent central, spin-spin, tensor and spin-orbit potentials, 
 respectively.
 
 In this paper, we focus only on the $S$-wave charmonium 
 states ($\eta_c$ and $J/\psi$).  We perform an appropriate projection 
 with respect to the discrete rotation, which provides 
 the BS wave function projected in the $A^{+}_{1}$ representation,  
 $\phi_{\Gamma}({\bf r})\rightarrow \phi_{\Gamma}(A^{+}_1; r)$.
 This projected BS wave function corresponds to the $S$-wave 
 in continuum theory at low energy~\cite{Luscher:1990ux}. 
 We simply denote the $A_1^{+}$ projected
 BS wave function by $\phi_{\Gamma}(r)$ hereafter.
 
 The stationary Schr\"odinger equation for the projected 
 BS wave function $\phi_{\Gamma}(r)$ is reduced to
 \begin{equation}
   \left\{
   - \frac{\nabla^2}{\mQ}
   +V(r)+\SdotS V_{\text{S}}(r)
   \right\}\phi_{\Gamma}(r)=E_\Gamma \phi_{\Gamma}(r)
   \label{Eq_pot}
 \end{equation}
 at the leading order of the $v$-expansion~\cite{footnote2}.
 The spin operator $\SdotS$ can be easily replaced by expectation values 
 $-3/4$ and $1/4$ for the PS and V channels, respectively. 
 As a result, both spin-independent and -dependent part of 
 the central interquark potentials can be separately evaluated 
 through a linear combination of Eq.(\ref{Eq_pot}) calculated 
 for both PS and V channels as
 \begin{eqnarray}
   V(r)
   &=& E_{\text{ave}}+\frac{1}{\mQ}\left\{
   \frac{3}{4}\frac{\nabla^2\phi_\text{V}(r)}{\phi_\text{V}(r)}+
    \frac{1}{4}\frac{\nabla^2\phi_\text{PS}(r)}{\phi_\text{PS}(r)}
   \right\} \label{Eq_potC}\\
   V_{\text{S}}(r) 
   &=& E_{\text{hyp}} + \frac{1}{\mQ}\left\{
  \frac{\nabla^2\phi_\text{V}(r)}{\phi_\text{V}(r)} 
  - \frac{\nabla^2\phi_\text{PS}(r)}{\phi_\text{PS}(r)} \right\},\label{Eq_potS}
 \end{eqnarray}
 where $E_{\text{ave}}=M_{\text{ave}}-2\mQ$ 
 and $E_{\text{hyp}}=M_\text{V}-M_\text{PS}$.
 The mass $M_{\text{ave}}$ denotes the spin-averaged mass as  
 $\frac{1}{4}M_\text{PS}+\frac{3}{4}M_\text{V}$.
 The derivative $\nabla^2$ is defined by the discrete Laplacian 
 with nearest-neighbor points.
 As for other spin-dependent potentials such as the tensor potential 
 $V_{\text{T}}$ and the spin-orbit potential $V_{\text{LS}}$, 
 in principle, this approach can access them by considering 
 the $P$-wave quarkonium states such as the $\chi_{c}$ ($0^{++}$, $1^{++}$) 
 and $h_c$ ($1^{+-}$) states, which must leave contributions 
 of $V_{\text{T}}$ and $V_{\text{LS}}$ to Eq.(\ref{Eq_pot}).
 
 It is worth mentioning that the quark kinetic mass $\mQ$, which 
 is essentially involved in the definition of the interquark potentials, 
 can be self-consistently evaluated within 
 the same framework~\cite{Kawanai:2011xb}. 
 The proper determination of the quark mass has a key role 
 in establishing the connection to the static heavy-quark potential 
 given by Wilson loops in the infinitely heavy-quark limit. 
 A more detailed discussion can be found in Ref.~\cite{Kawanai:2011xb}.
 
 To calculate the charmonium potential, we have performed 
 dynamical lattice QCD simulations on a lattice $L^3\times T=32^3\times 64$ 
 with 2+1 flavor PACS-CS gauge configurations generated 
 by Iwasaki gauge action at $\beta=1.90$, 
 which corresponds to a lattice cutoff of $a^{-1}\approx 2.2$ GeV 
 ($a \approx 0.091$fm)~\cite{Aoki:2008sm}.
 The spatial lattice size corresponds to $L \approx 3\;{\rm fm}$.
 The hopping parameters for the light sea quarks 
 \{$\kappa_{ud}$,$\kappa_{s}$\}=\{0.13781, 0.13640\}
 correspond to $M_\pi=156(7)$ MeV and $M_K= 554(2)$ MeV~\cite{Aoki:2008sm}.
 Although the light sea quark masses are slightly off the physical point, 
 the systematic uncertainty due to this fact could be extremely small 
 in our project. Our results are analyzed on all 198 gauge configurations, 
 which are available through International Lattice Data Grid and 
 Japan Lattice Data Grid~\cite{ILDG}.
 We fix gauge configurations to Coulomb gauge.
 
 For the charm quark, we employ the RHQ action to control  systematic
 uncertainties coming from the discretization 
 error induced by large quark mass~\cite{Aoki:2001ra}.
 The RHQ action utilized here is a variant of the Fermilab
 approach~\cite{ElKhadra:1996mp} and has five parameters $\kappa_c$,
 $\nu$, $r_s$, $c_B$ and $c_E$. %(See Ref.~\cite{Aoki:2001ra} for details).
 The parameters $r_s$, $c_B$ and $c_E$ are determined 
 by tadpole improved one-loop perturbation theory~\cite{Kayaba:2006cg}. 
 For $\nu$, we use a nonperturbatively determined value, 
 which is adjusted by reproducing the effective speed of light 
 to be unity in the dispersion relation 
 $E^2({\bf p}^2)=M^2+c^2_{\text{eff}}|{\bf p}|^2$ 
 for the spin-averaged $1S$ charmonium state, since the parameter $\nu$ 
 is sensitive to the size of hyperfine mass splitting~\cite{Namekawa:2011wt}.
 We choose $\kappa_c$ to reproduce the experimental spin-averaged mass
 of $1S$ charmonium states $M_\text{ave}^{\text{exp}}=3.0678(3)$ GeV.   
 To calibrate adequate RHQ parameters, we employ a gauge invariant 
 Gauss smearing source for the standard two-point correlation function
 with four different finite momenta. 
 As a result, the relevant speed of light, $c^2_{\text{eff}}=1.04(5)$, is
 observed for the spin-averaged mass of $1S$ charmonium states
 with our chosen RHQ parameters summarized in Table~\ref{Tab2}.
 
 We have computed charm quark propagators with two wall sources located 
 at different time slices $\tsrc/a=6$ and 57, to increase statistics.  
 Dirichlet boundary conditions are imposed for the time direction
 to eliminate unwanted contributions across time boundaries.
 We calculate a pair of four-point correlation functions 
 from two wall-source quark propagators and fold them together 
 to create a single four-point correlation function.
 
 Low-lying $S$- and $P$-wave charmonium masses measured in this study
 are all close to experimental values, though
 the hyperfine mass splitting $M_\text{hyp}=0.1124(9)$ GeV
 is slightly smaller than the experimental value, 
 $M_\text{hyp}^{\text{exp}}=0.1166(12)$ GeV.
   Note that we simply neglect the disconnected diagrams in both the $\eta_c$ and
   $J/\psi$ correlation functions.
 The similar value of the hyperfine mass splitting is reported even 
 on the physical point in Ref.~\cite{Namekawa:2011wt}.
 We summarize resulting charmonium masses in Table~\ref{Tab2}. 
 
 %
 % Table 2
 %  
 %\begin group
 %\squeezetable
 \begin{table}[th]
   \caption{Summary of the Cornell parameters and the quark mass
     determined from lattice QCD. For comparison, the corresponding values 
     adopted in a NRp model~\cite{Barnes:2005pb} 
     are also included.
     \label{Tab_para}
   }
   \begin{ruledtabular}
     \begin{tabular}{cccccc} %\hline  \hline 
       & \multicolumn{2}{c}{This work}& Polyakov lines & NRp model \\ 
       & on-axis & full set  & & \\
       \hline
       $A$           & 0.861(17)  & 0.813(22)  & 0.403(24) &0.7281 \\
       $\sqrt{\sigma}$ [GeV] & 0.394(7) & 0.394(7)   & 0.462(4) &0.3775 \\
       $\mQ$ [GeV]          & \multicolumn{2}{c}{1.74(3)}    &  $\infty$       &1.4794 \\
     \end{tabular}
   \end{ruledtabular}
 \end{table}
 %\endgroup
 
 %%%%%%%%%%%%%%%%%%%%%%%%%%%%%%%%%%%%%%%%%%%%%%%%%%%%%%
 % numerical results
 %%%%%%%%%%%%%%%%%%%%%%%%%%%%%%%%%%%%%%%%%%%%%%%%%%%%%%
 %
 %Results1
 %
 %
 %
 \begin{figure}
   \centering
   \includegraphics[width=.49\textwidth]{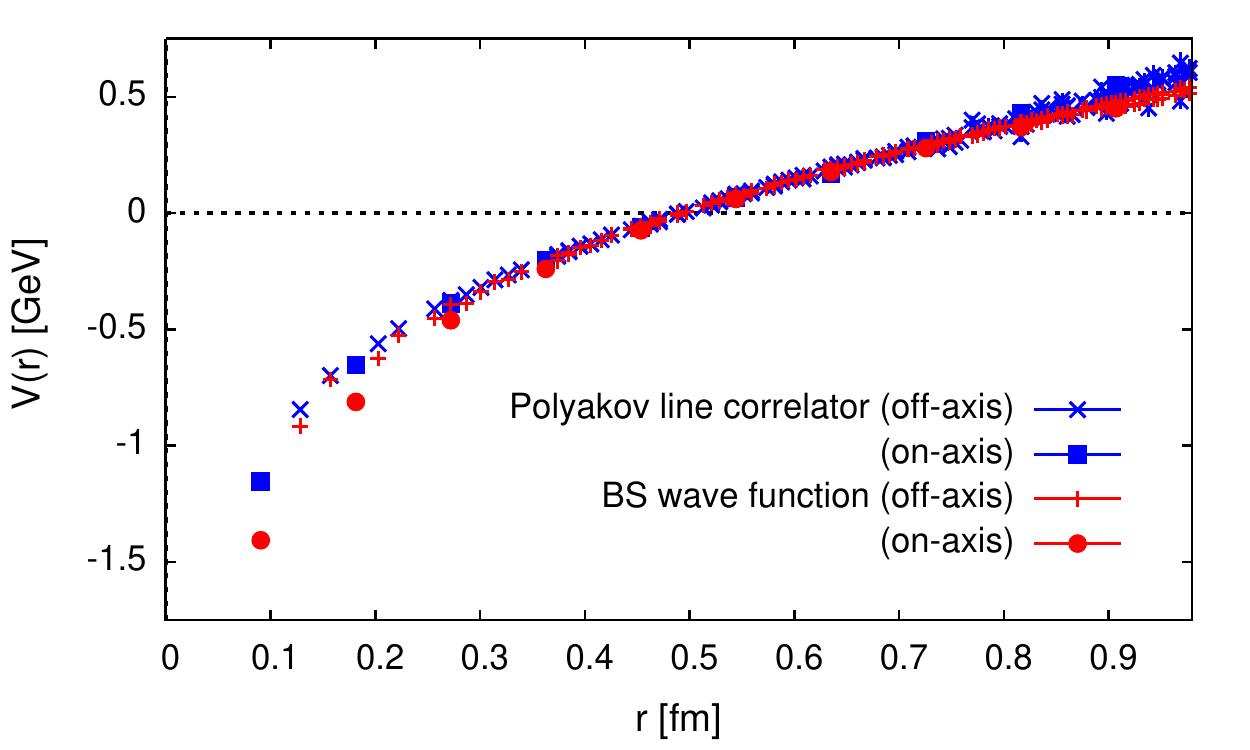}
   \caption{
     Spin-independent interquark potentials calculated from
     both the BS wave function of the charmonium states 
     ($\scriptstyle{+}$ and $\bullet$) and the standard 
     Polyakov line correlator ($\times$ and $\scriptstyle{\blacksquare}$).
     Filled symbols represent on-axis data. The constant terms are 
     subtracted from both potentials to set to $V(r_0)=0$.
     \label{Pot}}
 \end{figure}
 First, we show a result of the spin-independent charmonium potential $V(r)$
 in Fig.~\ref{Pot}, where the constant term is subtracted 
 to set $V(r_0)=0$ with the Sommer scale, $r_0\approx 0.5$~fm. 
 For comparison, the static heavy-quark potential calculated 
 from the Polyakov line correlator, which corresponds to the one
 obtained in the infinitely heavy-quark limit similar to the Wilson 
 loop results~\cite{Bali:2000gf}, is also displayed in Fig.~\ref{Pot}.
 As expected, the charmonium potential calculated in the BS amplitude 
 method properly exhibits the linearly rising potential 
 at large distances and the Coulomb-like potential at short distances. 
 
 Here we give some technical remarks on systematic uncertainties.
 In the BS amplitude method, we take a weighted average of 
 data points in the wide range of $|t-\tsrc|/a=26\text{\ -\ }48$
 for determining the equal-time BS wave function. Therefore, 
 the resulting charmonium potential has a much smaller systematic 
 error stemming from the uncertainty in the choice of time window
 than the conventional approach to calculate the static heavy-quark 
 potential by Wilson-loops or Polyakov lines.  On the other hand, 
 the discretization error seems to much severely appear 
 in the charmonium potential especially near the origin.
 The Coulomb-like behavior obtained in the BS amplitude method may contain
 large uncertainties, which should vanish in the continuum limit. 
 To avoid the large discretization error, we hereafter prefer 
 to use the ``on-axis'' data, which less suffers from the rotational 
 symmetry breaking in the finite cubic box.
 
 %
 %Results2
 %
 %
 %
 \begin{figure}
   \centering
   \includegraphics[width=.49\textwidth]{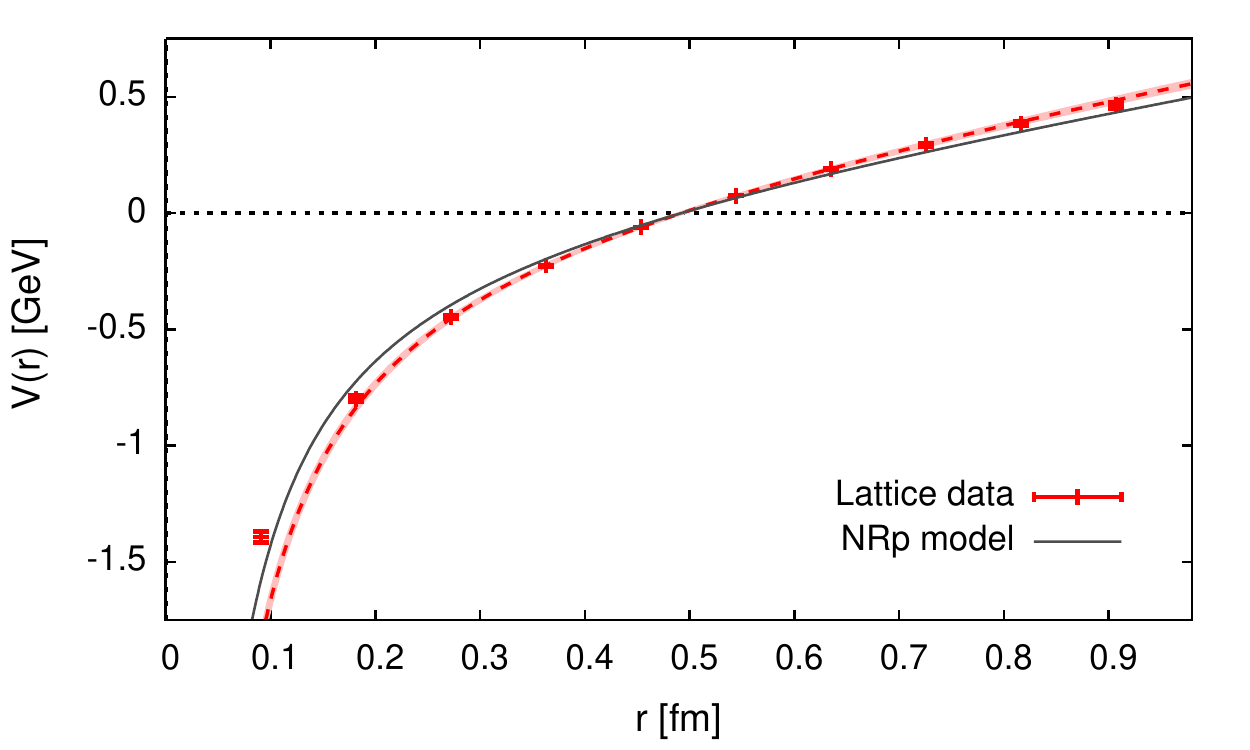}
   \caption{Spin-independent charmonium potential calculated from the
     BS wave function. The dashed curve is the fitting result
     by the Cornell parametrization. The shaded band shows
     statistical fitting uncertainty calculated by the jack-knife method.
     For comparison, the phenomenological potential adopted in a
     NRp model~\cite{Barnes:2005pb} is also included as
     solid curve.
     \label{Pot2}}
 \end{figure}

 The Cornell parametrization is simply adopted 
 for fitting our data of the spin-independent central potential:
  \begin{equation} 
    V(r)= -\frac{A}{r}+\sigma r + V_0 \label{Eq_Cornell}
  \end{equation}
  with the Coulombic coefficient $A$, the string tension $\sigma$ 
  and a constant $V_0$. We have carried out correlated $\chi^2$ fits 
  with full covariance matrix for on-axis data over range 
  $4\leq r/a \leq 10$, while uncorrelated fits are adopted 
  in full data analysis including all off-axis data points due to 
  high correlation between different $r$ points.
  The fitting results are listed in Table~\ref{Tab2} 
  together with the phenomenological values employed 
  by a nonrelativistic potential~(NRp) model in Ref.~\cite{Barnes:2005pb}.
  From on-axis data only, we get the Cornell parameters of 
  the charmonium potential: $A=0.861(17)$ and $\sqrt{\sigma}=0.394(7)$ MeV 
  with acceptable $\chi^2/\text{dof}$ ($\approx 2.2$).
  The quoted errors represent only the statistical errors
  given by the jack-knife analysis.
 
  In Fig.~\ref{Pot2}, we show on-axis data points of the spin-independent 
  charmonium potential with the fitted curve (dashed curve).
  The phenomenological potential used in NRp models~\cite{Barnes:2005pb}  
  is also plotted as a solid curve for comparison.
  As shown in Fig.~\ref{Pot2}, although the charmonium potential obtained from 
  lattice QCD is quite similar to the one in the NRp models,
  the string tension of the charmonium potential is slightly stronger 
  than the phenomenological one.  Therefore our result indicates that 
  there are only minor modifications required for the spin-independent 
  central potential in the NRp models.
  
 Moreover, it seems that a gap for the Coulombic coefficients between 
  the conventional static potential from Wilson-loops and
  the phenomenological potential used in the NRp models closes by 
  our new approach, which nonperturbatively accounts for a finite quark mass effect.  
  
  It is worth mentioning that the string breaking, which would be 
  induced by the presence of dynamical quarks was not observed 
  at least in the range $r\alt 1$ fm, where we still get a better 
  signal-to-noise ratio. It is indeed  what we expected, since
  we cannot access information of the potential outside of 
  the localized wave function, which represents the charmonium bound state 
  within the BS amplitude method.  We here calculate only the BS wave 
  functions of $1S$ charmonium states, which are quickly dumped around 
  outside of $r\agt 1$ fm. Therefore, at least the similar calculation for 
  the higher-lying charmonium states, whose wave function
  can be extended until the string breaking sets in, is demanded to observe 
  such effect.
  
  In this calculation, the kinetic mass of the charm quark is determined
  self-consistently within the BS amplitude method as well. 
  (See Ref.~\cite{Kawanai:2011xb} for details.) 
  The charm quark mass obtained in this study is about $17\%$ heavier 
  than the one adopted in the NRp models, of which value is also listed 
  in Table~\ref{Tab_para}. 
  This difference should not be taken seriously 
  since the spatial profile of the spin-spin potential from lattice QCD 
  is slightly different from the one used in the NRp models 
  as we will discuss later.  

  %
  %Results2
  %
  %
  \begin{figure}
    \centering
    \includegraphics[width=.49\textwidth]{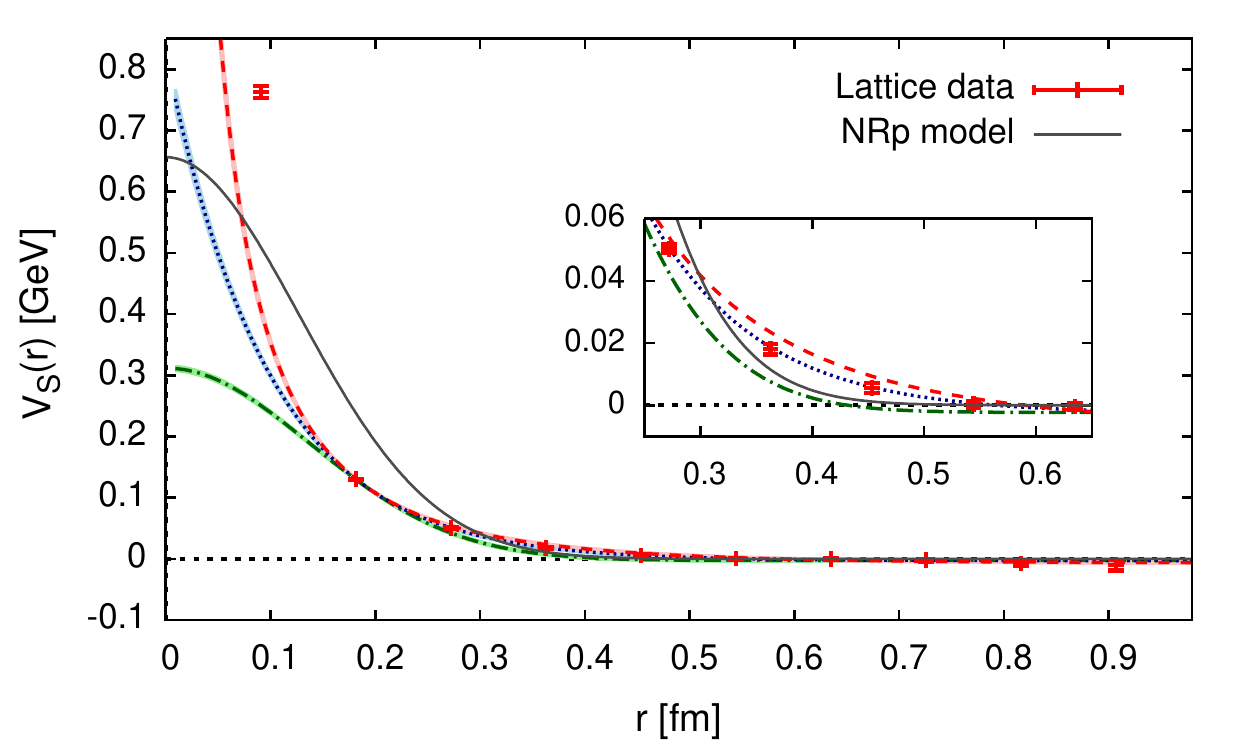}
    \caption{
      Spin-spin charmonium potential calculated from the BS wave
      function. The dashed, dotted and dash-dotted curve
      correspond to fitting results of Yukawa, exponential and
      Gaussian functional forms, respectively.
      For comparison, the phenomenological potential adopted in a NRp
      model~\cite{Barnes:2005pb} is also included as solid curve.
      \label{Pot_hyp}}
  \end{figure}
  In Fig.~\ref{Pot_hyp}, we show the spin-spin term of 
  the charmonium potential and the corresponding phenomenological one 
  found in Ref.~\cite{Barnes:2005pb}. Our spin-spin potential exhibits 
  the short-range {\it repulsive interaction}, 
  which is required by the charmonium spectroscopy, where the 
  higher spin state in hyperfine multiplets receives heavier mass.
  It should be reminded that the Wilson loop approach
  fails to reproduce the correct behavior of the spin-spin interaction, 
  since the leading-order spin-spin potential classified in potential nonrelativistic QCD
  becomes attractive at short distances~\cite{Koma:2006fw}.
  
  %
  % Table
  %  
  %\begingroup
  %\squeezetable
  \begin{table}%[H]
    \caption{Results of fitted parameters for the spin-spin potential
      with three types of the fitting functional form.
      \label{Tab_hyp}
    }
    \begin{ruledtabular}
      \begin{tabular}{lllc} %\hline  \hline 
        functional form& \ \ \ \ \ \ $\alpha$ 
        & \ \ \ \ \ \ $\beta$ & $\chi^2/\text{dof}$\\ \hline
        Yukawa-type & 0.287(8) & 0.894(32) GeV & 7.28\\
        Exponential-type  & 0.825(19) GeV & 1.982(24) GeV& 1.46\\
        Gaussian-type & 0.314(4) GeV & 1.020(11) GeV$^2$ &22.79
      \end{tabular}
    \end{ruledtabular}
  \end{table}
  %\endgroup
  In contrast of the case of the spin-independent potential, 
  the spin-spin potential obtained here is absolutely different 
  from a repulsive $\delta$-function potential generated 
  by perturbative one-gluon exchange, which is widely adopted 
  in the NRp models.  However, such the contact form 
  $\propto \delta({\bf r})$ of the Fermi-Breit type potential 
  is not reliable since the pointlike spin-spin interaction can 
  not give a finite hyperfine mass splitting of the $P$- or 
  higher-wave charmonia~\cite{Voloshin:2007dx}.
  Indeed, the finite-range spin-spin potential described 
  by the Gaussian form is adopted in Ref.~\cite{Barnes:2005pb}, where 
  many properties of conventional charmonium states at higher masses 
  are predicted. 
  
  This phenomenological spin-spin potential is also plotted 
  in Fig.~\ref{Pot_hyp} for comparison. 
  Although there is a slight 
  difference at very short distances, where 
  systematic uncertainties become severe
  due to lattice discretization errors,
  the spin-spin potential from first principles QCD 
  is barely consistent with the phenomenological one.    
    
  To examine the appropriate functional form for the spin-spin potential, 
  we have tried three types of functional forms: 
  \begin{equation}
    V_{\text{S}}(r)=\left\{
    \begin{array}{lcl}
      \alpha \exp(-\beta r)/r &:& \text{Yukawa form} \\
      \alpha \exp(-\beta r)   &:& \text{Exponential form} \\
      \alpha \exp(-\beta r^2) &:& \text{Gaussian form}.
    \end{array}
    \right.
  \end{equation}
  We then determine which functional form can give a reasonable fit 
  over the range of $r/a$ from 2 to 10.  All results of correlated 
  $\chi^2$ fits are summarized in Table~\ref{Tab_hyp}.
  The long-range screening observed in the spin-spin potential
  is more easily accommodated by the Yukawa form or the exponential form
  than the Gaussian form that is often employed in the NRp models.
  Although the exponential form provides the smaller $\chi^2/\text{dof}$ than 
  the Yukawa form, a solid conclusion requires more accurate information on 
  the short-range behavior of the spin-spin potential.  
  
  %
  %summary
  %
  In this paper, we have studied both spin-independent and spin-dependent part
  of the charmonium potential by means of the BS wave function of 
  $1S$ charmonium  states in dynamical lattice QCD simulations.
  The spin-independent charmonium potential obtained from lattice QCD 
  with almost physical quark masses is quite similar to  the one used 
  in the NRp models. 
   The spin-spin potential, which is, for the first time, 
  determined in dynamical lattice simulations, 
  exhibits not pointlike, but finite-range repulsive interaction. Its $r$-dependence is barely consistent with  
  the phenomenological one adopted in Ref.~\cite{Barnes:2005pb}. 
  Thus, a full set of the reliable spin-dependent 
  potentials derived from lattice QCD within our approach can provide new and valuable 
  information to the NRp models. This improvement of the spin-dependent
  potentials will help in making accurate 
  theoretical predictions about the higher-mass charmonium states. 
  We plan to extend our research to determine all spin-dependent terms 
  in the charmonium potential, including the tensor and spin-orbit forces 
  and also to address all the possible systematic uncertainties
  described in the text.

 %
 %acknowledgement
 %
  We acknowledge the PACS-CS collaboration and ILDG/JLDG~\cite{ILDG} for 
  providing us with the gauge configurations. We would also like to
  thank H. Iida, Y. Ikeda and T. Hatsuda for fruitful discussions.
  This work was partially supported by JSPS/MEXT Grants-in-Aid
  (No.~22-7653, No.~19540265, No.~21105504 and No.~23540284).


\begin{references}
    
    %\cite{Voloshin:2007dx}
  \bibitem{Voloshin:2007dx}
    M.~B.~Voloshin,
    %``Charmonium,''
  Prog.\ Part.\ Nucl.\ Phys.\  {\bf 61}, 455 (2008).
  %[arXiv:0711.4556 [hep-ph]].
  %%CITATION = PPNPD,61,455;%%
  
  %\cite{Godfrey:2008nc}
\bibitem{Godfrey:2008nc}
  S.~Godfrey and S.~L.~Olsen,
  %``The Exotic XYZ Charmonium-like Mesons,''
  Ann.\ Rev.\ Nucl.\ Part.\ Sci.\  {\bf 58}, 51 (2008).
  %[arXiv:0801.3867 [hep-ph]].
  %%CITATION = ARNUA,58,51;%%
  
   %\cite{Eichten:1975ag}
  \bibitem{Eichten:1975ag}
  E.~Eichten {\it et.al.}, %K.~Gottfried, T.~Kinoshita, K.~D.~Lane and T.~M.~Yan,
  %``The Interplay Of Confinement And Decay In The Spectrum Of Charmonium,''
  Phys.\ Rev.\ Lett.\  {\bf 36}, 500 (1976).
  %%CITATION = PRLTA,36,500;%%

   %\cite{Godfrey:1985xj}
  \bibitem{Godfrey:1985xj}
  S.~Godfrey and N.~Isgur,
  %``Mesons In A Relativized Quark Model With Chromodynamics,''
  Phys.\ Rev.\  D {\bf 32}, 189 (1985).
  %%CITATION = PHRVA,D32,189;%%
  
    %\cite{Barnes:2005pb}
\bibitem{Barnes:2005pb}
  T.~Barnes, S.~Godfrey and E.~S.~Swanson,
  %``Higher Charmonia,''
  Phys.\ Rev.\  D {\bf 72}, 054026 (2005).
  %[arXiv:hep-ph/0505002].
  %%CITATION = PHRVA,D72,054026;%%
%%%% pNRQCD 

  %\cite{Bali:2000gf}
  \bibitem{Bali:2000gf}
  For a review, see G.~S.~Bali,
  %``QCD forces and heavy quark bound states,''
  Phys.\ Rept.\  {\bf 343}, 1 (2001).
  %[arXiv:hep-ph/0001312].
  %%CITATION = PRPLC,343,1;%%

%\cite{Brambilla:2004jw}
\bibitem{Brambilla:2004jw}
  %For a review, see
  N.~Brambilla {\it et al.}, %A.~Pineda, J.~Soto and A.~Vairo,
  %``Effective field theories for heavy quarkonium,''
  Rev.\ Mod.\ Phys.\  {\bf 77}, 1423 (2005).
  %[arXiv:hep-ph/0410047].
  %%CITATION = RMPHA,77,1423;%%


  %\cite{Koma:2006fw}
  \bibitem{Koma:2006fw}
  Y.~Koma and M.~Koma,
  %``Spin-dependent potentials from lattice QCD,''
  Nucl.\ Phys.\  B {\bf 769}, 79 (2007).
  %[arXiv:hep-lat/0609078].
  %%CITATION = NUPHA,B769,79;%%

%\cite{Kawanai:2011xb}
\bibitem{Kawanai:2011xb}
  T.~Kawanai and S.~Sasaki,
  %``Interquark potential with finite quark mass from lattice QCD,''
  Phys.\ Rev.\ Lett.\  {\bf 107}, 091601 (2011).
  %[arXiv:1102.3246 [hep-lat]].
  %\cite{Kawanai:2011xb}

 %\cite{Ikeda:2011bs}
\bibitem{Ikeda:2011bs}
  Y.~Ikeda and H.~Iida,
  %``Quark-anti-quark potentials from Nambu-Bethe-Salpeter amplitudes on
  %lattice,''
  arXiv:1102.2097 [hep-lat].
  %%CITATION = ARXIV:1102.2097;%%

  %\cite{Ishii:2006ec}
  \bibitem{Ishii:2006ec}
  N.~Ishii, S.~Aoki and T.~Hatsuda,
  %``The nuclear force from lattice QCD,''
  Phys.\ Rev.\ Lett.\  {\bf 99}, 022001 (2007),
  %[arXiv:nucl-th/0611096].
  %%CITATION = PRLTA,99,022001;%%
  %\cite{Aoki:2009ji}
  %\bibitem{Aoki:2009ji}
  S.~Aoki, T.~Hatsuda and N.~Ishii,
  %``Theoretical Foundation of the Nuclear Force in QCD and its applications to
  %Central and Tensor Forces in Quenched Lattice QCD Simulations,''
  Prog.\ Theor.\ Phys.\  {\bf 123} (2010) 89.
  %[arXiv:0909.5585 [hep-lat]].
  %%CITATION = PTPKA,123,89;%%

%\cite{Aoki:2008sm}
\bibitem{Aoki:2008sm}
  S.~Aoki {\it et al.}  [PACS-CS Collaboration],
  %``2+1 Flavor Lattice QCD toward the Physical Point,''
  Phys.\ Rev.\  D {\bf 79}, 034503 (2009).
  %[arXiv:0807.1661 [hep-lat]].
  %%CITATION = PHRVA,D79,034503;%%

    %\cite{Aoki:2001ra}
  \bibitem{Aoki:2001ra}
  S.~Aoki, Y.~Kuramashi and S.~I.~Tominaga,
  %``Relativistic heavy quarks on the lattice,''
  Prog.\ Theor.\ Phys.\  {\bf 109}, 383 (2003).
  %[arXiv:hep-lat/0107009].
  %%CITATION = PTPKA,109,383;%%

%%% qqbar BS amplitude %%%%
  %\cite{Velikson:1984qw}
  \bibitem{Velikson:1984qw}
  B.~Velikson and D.~Weingarten,
  %``Hadron Wave Functions In Lattice QCD,''
  Nucl.\ Phys.\  B {\bf 249}, 433 (1985).
  %%CITATION = NUPHA,B249,433;%%
  %\cite{Gupta:1993vp}

  \bibitem{Gupta:1993vp}
  R.~Gupta, D.~Daniel and J.~Grandy,
  %``Bethe-Salpeter amplitudes and density correlations for mesons with Wilson
  %fermions,''
  Phys.\ Rev.\  D {\bf 48}, 3330 (1993).
  %[arXiv:hep-lat/9304009].
  %%CITATION = PHRVA,D48,3330;%%

%\cite{Caswell:1978mt}
\bibitem{Caswell:1978mt} 
  W.~E.~Caswell and G.~P.~Lepage,
  %``Reduction Of The Bethe-salpeter Equation To An Equivalent Schrodinger Equation, With Applications,''
  Phys.\ Rev.\ A {\bf 18}, 810 (1978).
  %%CITATION = PHRVA,A18,810;%%
  
\bibitem{footnote1} 
The relativistic effect has been estimated using 
relativistic kinematics in Ref.~\cite{Ikeda:2011bs}.
Although the short-range behavior of interquark potential 
is slightly influenced by this modification,
it is indeed small for the heavier quarks such as the charm quark.


  
  %\cite{Luscher:1990ux}
  \bibitem{Luscher:1990ux}
  M.~L\"uscher,
  %``Two Particle States On A Torus And Their Relation To The Scattering
  %Matrix,''
  Nucl.\ Phys.\ B {\bf 354}, 531 (1991).
  %%CITATION = NUPHA,B354,531;%%    
  
   \bibitem{footnote2}
  Here, we essentially follow the NRp models,
   where the $J/\psi$ state is purely composed of the $1S$ wave function. 
   However, within this method, this assumption can be verified 
   by evaluating the size of a mixing between $1S$ and $1D$ wave
   functions in principle.

  \bibitem{ILDG}
  International Lattice Data Grid/Japan Lattice Data Grid,
  http://www.jldg.org.
 
  %\cite{ElKhadra:1996mp}
  \bibitem{ElKhadra:1996mp}
 A.~X.~El-Khadra, A.~S.~Kronfeld and P.~B.~Mackenzie,
  %A.~X.~El-Khadra {\it et al.}, %, A.~S.~Kronfeld and P.~B.~Mackenzie,
  %``Massive Fermions in Lattice Gauge Theory,''
  Phys.\ Rev.\  D {\bf 55}, 3933 (1997).
  %[arXiv:hep-lat/9604004].
  %%CITATION = PHRVA,D55,3933;%%

  
  %\cite{Kayaba:2006cg}
  \bibitem{Kayaba:2006cg}
  Y.~Kayaba {\it et al.}  [CP-PACS Collaboration],
  %``First Nonperturbative Test of a Relativistic Heavy Quark Action in Quenched
  %Lattice QCD,''
  JHEP {\bf 0702}, 019 (2007).
  %[arXiv:hep-lat/0611033].
  %%CITATION = JHEPA,0702,019;%%
  

  %\cite{McNeile:2004wu}
  %\bibitem{McNeile:2004wu}
  %C.~McNeile and C.~Michael  [UKQCD Collaboration],
  %``An Estimate of the flavor singlet contributions to the hyperfine splitting
  %in charmonium,''
%  Phys.\ Rev.\  D {\bf 70}, 034506 (2004)
  %[arXiv:hep-lat/0402012].
  %%CITATION = PHRVA,D70,034506;%%

  %\cite{Namekawa:2011wt}
  \bibitem{Namekawa:2011wt}
  Y.~Namekawa {\it et al.}  [PACS-CS Collaboration],
  %``Charm quark system at the physical point of 2+1 flavor lattice QCD,''
  arXiv:1104.4600 [hep-lat].
  %%CITATION = ARXIV:1104.4600;%%

 \end{references}
\end{document}